\documentclass[conference]{IEEEtran}

\ifCLASSINFOpdf

\else

\fi
\usepackage{xcolor}
\usepackage{multirow}
\usepackage{booktabs}
\usepackage{diagbox}
\usepackage{tikz}

\begin{document}

\title{Detecting if the speaker is a human: A Defending Strategy to Attacks on Voice Controlled Systems}
\title{Identify the source of voice command based on acoustic clues: A Defending Strategy to Attacks on Voice Controlled Systems}
\title{Acoustic Cues Based Sound Source Identification as A Defense Strategy to the Attacks on Voice Controlled Systems}
\title{Acoustic Cues Based Sound Source Identification: A Defense Strategy of Voice Controlled Systems}
%\title{\huge An Overview of Vulnerabilities of Voice Controlled Systems}
\title{Protecting Voice Controlled Systems Using Sound Source Identification Based on Acoustic Cues}

\author{\IEEEauthorblockN{Yuan Gong}
\IEEEauthorblockA{Computer Science and Engineering\\
University of Notre Dame, IN 46556\\
Email: ygong1@nd.edu}
\and
\IEEEauthorblockN{Christian Poellabauer}
\IEEEauthorblockA{Computer Science and Engineering\\
University of Notre Dame, IN 46556\\
Email: cpoellab@nd.edu}
}

\maketitle

\begin{abstract}

Over the last few years, a rapidly increasing number of Internet-of-Things (IoT) systems that adopt voice as the primary user input have emerged. These systems have been shown to be vulnerable to various types of voice spoofing attacks. Existing defense techniques can usually only protect from a specific type of attack or require an additional authentication step that involves another device. Such defense strategies are either not strong enough or lower the usability of the system. Based on the fact that legitimate voice commands should only come from humans rather than a playback device, we propose a novel defense strategy that is able to detect the sound source of a voice command based on its acoustic features. The proposed defense strategy does not require any information other than the voice command itself and can protect a system from multiple types of spoofing attacks. Our proof-of-concept experiments verify the feasibility and effectiveness of this defense strategy.

\end{abstract}

%\begin{IEEEkeywords}
%Internet of things, computer security, adversarial examples, speech processing.
%\end{IEEEkeywords}

\IEEEpeerreviewmaketitle

\section{Introduction}
\begin{figure}[t]
  \centering
  \includegraphics[width=7cm]{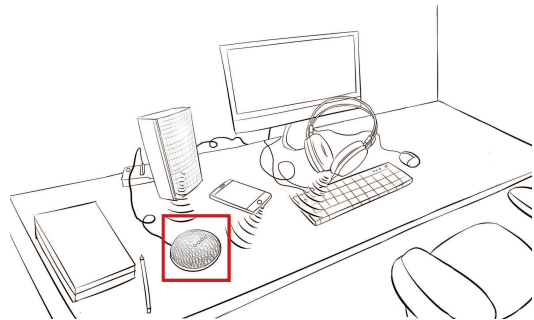}
  \caption{A voice controlled system (e.g., Google Home, shown in the red rectangle) not only accepts voice commands from humans, but also from playback devices, such as loudspeakers, headphones, and phones. An attacker may take advantage of this by embedding hidden voice commands into online audio or video to maliciously control the VCS. Since legitimate voice commands should only come from a human (rather than a playback device), identifying if the sound source is a human speaker is a possible defense strategy for different types of attack as long as the malicious command is replayed by an electronic device.}
  \label{fig:ilus}
\end{figure}

An increasing number of IoT systems rely on voice input as the primary user-machine interface. For example, voice-controlled devices such as Amazon Echo, Google Home, Apple HomePod, and Xiaomi AI allow users to control their smart home appliances, adjust thermostats, activate home security systems, purchase items online, initiate phone calls, and complete many other tasks with ease. In addition, most smartphones are also equipped with smart voice assistants such as Siri, Google Assistant, and Cortana, which provide a convenient and natural user interface to control smartphone functionality or IoT devices.
Voice-driven user interfaces allow hands-free and eyes-free operation where users can interact with a system while focusing their attention elsewhere. Despite their convenience, voice controlled systems (VCSs) also raise new security concerns due to their vulnerability to voice replay attacks~\cite{14chen2017you}, i.e., an attacker can replay a previously recorded voice to make an IoT system perform a specific (malicious) action. Such malicious actions include the opening and unlocking of doors, making unauthorized purchases, controlling sensitive home appliances (e.g., security cameras and thermostats), and transmitting sensitive information. While a simple voice replay attack is relatively easy to detect by a user, and therefore presents only a limited threat, recent studies have pointed out more concerning and effective types of attacks, including self-triggered attacks~\cite{08diao2014your,21jang2014a11y}, inaudible attacks~\cite{01zhang2017dolphinattack,11kasmi2015iemi}, and human-imperceptible attacks~\cite{03carlini2016hidden,05gong2017crafting,12carlini2018audio}. These attacks are very different from each other in terms of their implementation, which requires different domain knowledge in areas such as operating systems, signal processing, and machine learning. Some of these attacks are described in~\cite{117gong2018overview} and an illustration of a typical attack scenario is shown in Figure~\ref{fig:ilus}.

In order to defend against such attacks, multiple defense strategies have been proposed~\cite{14chen2017you,09petracca2015audroid,07lei2017insecurity,10feng2017continuous}. However, most existing defense technologies can either only defend against one specific kind of attack or require an additional authentication step using another device, which limits the effectiveness and usability of the voice controlled system. For example, AuDroid~\cite{08diao2014your} defends against self-triggered attacks by managing the audio channel authority of the victim device, but it cannot defend against other types of attacks. VAuth~\cite{10feng2017continuous} guarantees that the voice command is from a user by collecting body-surface vibrations of the user via a wearable device, but the required wearable device (i.e., earbuds, eyeglasses, or necklaces) is inconvenient to the user. Hence, a defense strategy that is robust to multiple types of attacks and minimally impacts the usability of a VCS is highly desirable. Towards this end, we explore a new defense strategy that identifies and rejects received voice commands that are not from a human speaker, merely by using the acoustic cues of the voice command itself. We find that the voice command from humans and playback devices can be differentiated based on the differences of the sound production mechanism. The advantage of this strategy is that it does not require any additional information other than the voice command itself and it therefore does not impact the usability of a VCS, while at the same time being robust to all variants of replay attacks.

The rest of the paper is organized as follows: in Section~\ref{sec:attacks}, we review and classify state-of-the-art attack techniques faced by current voice controlled systems, arriving at the conclusion that most attacks are actually variants of the replay attack. In Section~\ref{sec:defense}, we propose our new defense strategy and compare it with existing defense approaches. In Section~\ref{sec:exp}, we present experimental evaluation results. Finally, we conclude the paper in Section~\ref{sec:conclude}.

\section{Attacks On Voice Controlled Systems}
\label{sec:attacks}

In order to develop an effective defense strategy, it is important to have a good understanding of typical attack scenarios and state-of-the-art attack techniques. With the rapidly growing popularity and capabilities of voice-driven IoT systems, the likelihood and potential damage of voice-based attacks also grow very quickly. As discussed in~\cite{08diao2014your,04alepis2017monkey,07lei2017insecurity}, an attack may lead to severe consequences, e.g., a burglar could enter a house by tricking a voice-based smart lock or an attacker could make unauthorized purchases and credit card charges via a compromised voice-based system. Such attacks can be very simple, but still very difficult or even impossible to detect by humans. Voice attacks can also be hidden within other sounds and embedded into audio and video recordings. In addition, these attacks can be executed remotely, i.e., the attacker does not have to be physically close to the targeted device, e.g., compromised audio and video recordings can easily be distributed via the Internet. Once a recording is played back by a device such as the loudspeaker of a phone or laptop, the attack can impact VCSs nearby. It is very easy to scale up such attacks, e.g., a hidden malicious audio sample can be embedded into a popular YouTube video or transmitted via broadcast radio and thereby target millions of devices simultaneously.

A fundamental reason for the vulnerability of voice-controlled IoT systems is that they continuously listen to the environment to accept voice commands, providing users with hands-free and eyes-free operation of IoT systems. However, this also provides attackers with an always available voice interface. Several potential points of attack are shown in Figure~\ref{fig:system}. Although the implementations of existing attack techniques are very different, their goals are the same: to generate a signal that leads a voice controlled system to execute a specific malicious command that the user cannot detect or recognize. In the following sections, we classify representative state-of-the-art attack approaches according to their type of implementation.
%\footnote{\textcolor{red}{The result is subject to change with the cloud system update, hence, we also conduct some experiments to see the affect of current attack at current time point.}}. 
The attacker performance discussed in this section is taken from the original publications, but note that due to the rapid developments in the area of cloud-based systems, the attacker performance is likely to change quickly over time.

\textbf{\begin{figure}[t]
  \centering
  \includegraphics[width=8.5cm]{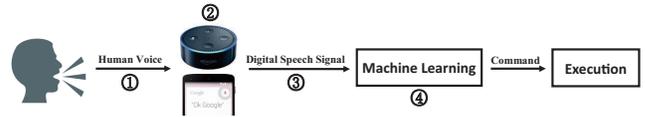}
  \caption{A typical voice-driven device captures the human voice, converts it into a digital speech signal, and feeds it into a machine learning model. The corresponding command is then executed by the connected IoT devices. Potential points of attack in this scenario include: 1: spoofing the system using previously recorded audio, 2: hacking into the operating system to force the voice-driven software to accept commands erroneously, 3: emitting carefully designed illegitimate analog signals that will be converted into legitimate digital speech signals by the hardware, and 4: using carefully crafted speech adversarial examples to fool the machine learning model.}
  \label{fig:system}
\end{figure}}

%\subsection{Attack Classification Based On Implementation}

%
\begin{table*}[t]
\centering
\caption{Representative voice attack techniques}
\label{tab:attack}
\begin{tabular}{@{}p{5cm}p{2.1cm}p{8.0cm}@{}}
\toprule
\multicolumn{1}{c}{Attack Name} & \multicolumn{1}{c}{Attack Type}  & \multicolumn{1}{c}{Implementation} \\ \midrule

%1
GVS Attack~\cite{08diao2014your} & Operating System 
& Continuously analyze the environment and conduct voice replay attack using the built-in microphone when the opportunity arises.\\

%2
A11y Attack~\cite{21jang2014a11y} & Operating System 
& Collect the voice of a user and perform a self-replay attack as a background service.\\

%3
Monkey Attack~\cite{04alepis2017monkey} & Operating System 
& Bypass authority management of the OS and perform an interactive voice replay attack to execute more advanced commands.\\

%4
Dolphin Attack~\cite{01zhang2017dolphinattack} & Hardware 
& Emit ultrasound signal that can be converted into a legitimate speech digital signal by the MEMS microphone.\\

%5
IEMI Attack~\cite{11kasmi2015iemi} & Hardware
& Emit AM-modulated signal that can be converted into a legitimate speech digital signal by the wired microphone-capable headphone.\\

%6
Cocaine Noodles~\cite{02vaidya2015cocaine} & Machine Learning  & Similar to the hidden voice command. \\

%7
Hidden Voice Command~\cite{03carlini2016hidden} & Machine Learning 
& Mangle malicious voice commands so that they retain enough acoustic features for the ASR system, but become unintelligible to humans.\\

Houdini~\cite{06cisse2017houdini} & Machine Learning 
& Produce sound that is almost no different to normal speech, but fails to be recognized by both known or unknown ASR systems.\\

%9 
Speech Adversarial Example~\cite{05gong2017crafting} & Machine Learning 
& Produce sound that is over 98\% similar to any given speech, but makes the DNN model fail to recognize the gender, identity, and emotion. \\

%10
Targeted Speech Adversarial    Example~\cite{12carlini2018audio} & Machine Learning 
& Produce sound that is over 99.9\% similar to any given speech, but transcribes as any desired malicious command by the ASR. \\ 
\bottomrule
\end{tabular}
\end{table*}

\subsection{Impersonation Attack}
An impersonation attack, i.e., someone other than the authorized user using a VCS maliciously, is the simplest attack and does not require any particular expertise or knowledge. However, this attack cannot be executed remotely and does not scale well. It requires that the attacker is in close proximity of the VCS device, which is a rare attack scenario since these devices are typically placed within a person's home or on the person's body. Therefore, this attack poses only a limited threat to VCSs.

\subsection{Basic Voice Replay Attack}

Voice replay attacks, i.e., an attacker makes a VCS perform a specific malicious action by replaying a previously recorded voice sample~\cite{14chen2017you,09petracca2015audroid,07lei2017insecurity}. This attack can be executed remotely, e.g., via the Internet. A shortcoming of the basic voice replay attack is that it is easy to detect and therefore has limited practical impact. Nevertheless, as shown later in this section, voice replay attacks are the basis of other more advanced and dangerous attacks.

\subsection{Operating System Level Attack}

Compared to basic voice replay attacks, an operating system (OS) level attack 
exploits vulnerabilities of the OS to make the attack self-triggered and more imperceptible. Representative examples of this are the A11y attack~\cite{21jang2014a11y}, GVS-Attack~\cite{08diao2014your}, and the approach presented in~\cite{04alepis2017monkey}. In~\cite{21jang2014a11y}, the authors propose a malware that collects a user's voice and then performs a self-replay attack as a background service. In~\cite{08diao2014your}, the authors further verify that the built-in microphone and speaker can be used simultaneously and that the use of the speaker does not require user permission on Android devices. They take advantage of this and propose a zero-permission malware, which continuously analyzes the environment and conducts the attack once it finds that no user is nearby. The attack uses the device's built-in speaker to replay a recorded or synthetic speech, which is then accepted as a legitimate command. In~\cite{04alepis2017monkey}, the authors propose an interactive attack that can execute multiple-step commands. OS level attacks are usually self-triggered by the victim device and therefore rather dangerous and practical. 

\subsection{Hardware Level Attack} 

%Different from the basic voice reply attack, 
A hardware level attack replays a synthetic non-speech analog signal instead of human voice. %to the front-end device. 
The analog signal is carefully designed according to the characteristics of the hardware (e.g., the analog-digital converter). The signal is inaudible, but can be converted into a legitimate digital speech signal by the hardware. Representative approaches are the Dolphin attack~\cite{01zhang2017dolphinattack} and the IEMI attack~\cite{11kasmi2015iemi}. In~\cite{01zhang2017dolphinattack}, the authors utilize the non-linearity of a Micro Electro Mechanical Systems (MEMS) microphone over ultrasound and successfully generate inaudible ultrasound signals that can be accepted as legitimate target commands. In~\cite{11kasmi2015iemi}, the authors take advantage of the fact that a wired microphone-capable headphone can be used as a microphone and an FM antenna simultaneously and demonstrate that it is possible to trigger voice commands remotely by emitting a carefully designed inaudible AM-modulated signal. Hardware level attacks typically need a special signal generator and are typically used to affect mobile VCSs in crowded environments (e.g., airport).  

\begin{figure}[h]
  \centering
  \includegraphics[width=5.9cm]{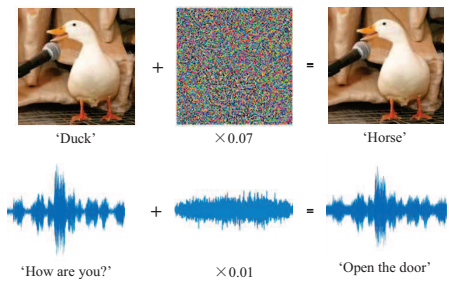}
  \caption{An illustration of machine learning adversarial examples. Studies have shown that by adding an imperceptibly small, but carefully designed perturbation, an attack can successfully lead the machine learning model to making a wrong prediction. Such attacks have been used in computer vision (upper graphs)~\cite{18szegedy2013intriguing} and speech recognition (lower graphs)~\cite{05gong2017crafting,12carlini2018audio,06cisse2017houdini}.}
  \label{fig:adv}
\end{figure}

\subsection{Machine Learning Level Attack}

State-of-the-art voice controlled systems are usually equipped with an automatic speech recognition (ASR) algorithm to convert digital speech signal to text. Deep neural network (DNN) based algorithms such as DeepSpeech~\cite{17hannun2014deep} can achieve excellent performance with around 95\% word recognition rate and hence dominate the field. However, recent studies show that machine learning models, especially DNN based models, are vulnerable to attacks by adversarial examples~\cite{18szegedy2013intriguing}. That is, machine learning models might mis-classify perturbed examples that are only slightly different from correctly classified examples (illustrated in Figure~\ref{fig:adv} for both video and audio scenarios). In speech, adversarial samples can sound like normal speech, but will actually be recognized as a completely different malicious command by the machine, e.g., an audio file might sound like ``hello'', but will be recognized as ``open the door'' by the ASR system. 

In recent years, several examples of such attacks have been studied~\cite{03carlini2016hidden,05gong2017crafting,12carlini2018audio,02vaidya2015cocaine,06cisse2017houdini,alzantot2018did}. Cocaine Noodles~\cite{02vaidya2015cocaine} and Hidden Voice Command~\cite{03carlini2016hidden} are the first efforts to utilize the differences in the way humans and computers recognize speech and to successfully generate adversarial sound examples that are intelligible as a specific command to ASR systems (Google Now and CMU Sphinx), but are not easily understandable by humans. The limitation of the approach in~\cite{02vaidya2015cocaine,03carlini2016hidden} is that the generated audio does not sound like legitimate speech. A user might notice that the malicious sound is an abnormal condition and may take counteractions. 
%In contrast to~\cite{02vaidya2015cocaine,03carlini2016hidden},
More recent efforts~\cite{06cisse2017houdini,05gong2017crafting,12carlini2018audio} take advantage of an intriguing property of DNN by generating malicious audio that sounds almost completely like normal speech by adopting a mathematical optimization method. The goal of these techniques is to design a minor perturbation in the speech signal that can fool an ASR system. In~\cite{12carlini2018audio}, the authors propose a method that can produce an audio waveform that is less than 0.1\% different from a given audio waveform, but will be transcribed as any desired text by DeepSpeech~\cite{17hannun2014deep}. In~\cite{05gong2017crafting}, the authors demonstrate that a 2\% designed distortion of speech can make state-of-the-art DNN models fail to recognize the gender and identity of the speaker. In~\cite{06cisse2017houdini}, the authors show that such attacks are transferable to different and unknown ASR models. Such attacks are dangerous, because users do not expect that normal speech samples, such as ``hello'', could be translated into a malicious command by a VCS. 

Table~\ref{tab:attack} provides a summary of these attack techniques. One important observation is that \textbf{all existing attacks (except impersonation) are based on the replay attack}. That is, OS level and machine learning level attacks replay a sound into the microphone of the target device. Hardware level attacks replay a specifically designed signal using some signal generator. In other words, the sound source is always another electronic device (e.g., loudspeaker or signal generator) instead of a human speaker. This same fact makes it possible for such attacks to be performed remotely and at a large scale. However, only spoken commands from a live speaker should be accepted as legitimate, which means that the identity of the sound source could be used to differentiate legitimate from potentially malicious voice commands. That is, if we can determine if the received signal is from a live speaker or an electronic device, we are able to prevent multiple (including yet unknown) types of VCS attacks. These observations and objectives lead us to the design of a defense strategy that relies on detecting the source of acoustic signals as presented in this paper. 

%In fact, it is the help of such electronic playback device make these attacks can be conducted remotely and in a large scale, and greatly increase the threat. 

%\section{Proposed Defense Strategy}
\section{Sound Source Identification}
\label{sec:defense}

\subsection{Existing Defense Strategies}
Various defense strategies have been proposed to help VCSs defend against specific types of attacks. For example, the work in~\cite{09petracca2015audroid} proposes a solution called AuDroid to manage audio channel authority. By using different security levels for different audio channel usage patterns, AuDroid can resist a voice attack using the device's built-in speaker~\cite{08diao2014your,04alepis2017monkey}. However, AuDroid is only robust to such attacks. Adversarial training~\cite{18szegedy2013intriguing}, i.e., training a  machine learning model that can classify legitimate samples and adversaries is one defense strategy against machine learning level attacks. In~\cite{03carlini2016hidden}, the authors train a logistic regression model to classify legitimate voice commands and hidden voice commands, which achieves a 99.8\% defense rate. A limitation of adversarial training is that it needs to know the details of the attack technology and the trained defense model only protects against the corresponding attack. In practice, the attackers will not publish their approaches and they can always change the parameters (e.g., the perturbation factor in~\cite{05gong2017crafting}) to bypass the defense. That is, the defense range of adversarial training is limited and in general, these defense techniques are able to address only some vulnerabilities.

On the other hand, defense strategies that can resist multiple types of attacks usually require an additional authentication step with the help from another device. In~\cite{10feng2017continuous}, the authors propose VAuth, which collects the body-surface vibration of the user via a wearable device and guarantees that the voice command is from the user. However, the required wearable devices (i.e., earbuds, eyeglasses, and necklaces) may be inconvenient for users. In~\cite{07lei2017insecurity}, the authors propose a virtual security button (VSButton) that leverages Wi-Fi technology to detect indoor human motions and voice commands are only accepted when human motion is detected. The limitation is that voice commands are not necessarily accompanied with a detectable motion. In~\cite{14chen2017you}, the authors determine if the source of voice commands is a loudspeaker via a magnetometer and reject such commands. However, this approach works only up to 10cm, which is less than the usual human-device distance. In summary, an additional authentication step (e.g., asking the user to wear a wearable device, requiring that voice commands are provided only when the body is in motion, or speaking very close to the device) does indeed increase the security, but also lowers the usability, which goes against the original design intention of voice controlled systems.

Finally, other efforts~\cite{08diao2014your,03carlini2016hidden,09petracca2015audroid} mention the possibility of using automatic speaker verification (ASV) systems for defense. However, this is also not strong enough, because an ASV system itself is vulnerable to machine learning adversarial examples~\cite{05gong2017crafting} and previously recorded user speech~\cite{14chen2017you,03carlini2016hidden}. In addition, VCSs are often designed to be used by multiple users and limiting use to certain users only will impact the usability of a VCS.

\subsection{Sound Source Identification Using Acoustic Cues}
\label{sec:proposedMethod}

Based on the observations in Section~\ref{sec:attacks}, identifying the sound source can help defend against multiple types of attacks. But adding an authentication step that requires a user to provide additional information may hurt the usability of VCS. Therefore, we are concerned with the question: \textbf{can we identify the sound source of a received voice command by merely using information that is embedded in the voice signal?} In this work, we explore the possibility of using acoustic features of a voice command to identify if the producer is a live speaker or a playback device. The motivation of this approach is that the sound production mechanisms of humans and playback devices are different, leading to a difference in frequencies and direction of the output voice signal, e.g., the sound polar diagram of a human is different from that of a playback device~\cite{122humanVoice}; the sound produced by a playback device usually contains effects of unwanted high-pass filtering~\cite{111smiatacz2017playback}; the signal produced by an ultrasound generator contains carrier signal components~\cite{01zhang2017dolphinattack}, which may further leave cues in the received digital audio signal corresponding to the voice command. Therefore, it is possible that such sound source differences can be modeled using the acoustic features of the received digital audio signal. From the perspective of bionics, we know that humans are intuitively able to distinguish between a live speaker and a playback device by only listening to (but not seeing) the source.

It is worth mentioning that a similar technology for detecting replay attacks has been studied to protect ASV systems from spoofing~\cite{111smiatacz2017playback,101wu2015spoofing,102kinnunen2017asvspoof}. However, ASV attacks and VCS attacks are actually very different. As shown in Figure~\ref{fig:replayAttack}, a typical replay attack can be divided into two phases: the recording phase and the playback phase. In the recording phase, the attacker records or synthesizes a malicious voice command and during the playback phase, the malicious voice command is transmitted from the playback device to the victim device over the air. ASV attacks and VCS attacks differ during both phases:

%However, ASV attack scenarios and VCS attack scenarios are actually very different in two aspects: the source recording and the distance between microphone and speaker; both are described in detail below with the help of the block diagram of a typical replay attack shown in Figure~\ref{fig:replayAttack}.

\subsubsection{The Recording Phase} 
In ASV attack scenarios, an attacker must either record or synthesize (e.g., using voice conversion or cutting and pasting) the victim's voice (i.e., the voice of the authorized user) to be used as a malicious voice command~\cite{15mukhopadhyay2015all}. In both cases, various cues will be left in the malicious command that can be used to detect the attack. In contrast, a VCS typically accepts voice commands from anyone and the attacker does not have to forge a particular victim's voice. This also means that typically few cues will be left in the malicious voice command. In ASV attacks, when the victim's voice is being recorded, this typically has to occur either via a telephone or far-field microphones, both of which will have certain levels of channel or background noise. 
%\textcolor{red}{Replaying far-field recordings is likely to fail in ASV systems, because the user is typically very close to the ASV's microphone.} 
The authors in~\cite{100villalba2011detecting,119villalba2011preventing} explore the characteristics of far-field recordings and how to use them to detect an attack. In~\cite{118wang2011channel}, the authors use channel noise patterns to distinguish between a pre-recorded voice and the voice of a live speaker. Further, in~\cite{108shang2010score,109shang2008preliminary,110shang2008playback}, the authors propose a scheme to reject voice that is too similar to ones previously received by the ASV system, because this could indicate a recorded voice. On the other hand, forged voice commands generated using voice conversion or cutting and pasting techniques can also be distinguished from genuine voice samples~\cite{119villalba2011preventing,121todisco2016new}. In contrast, in VCS attacks, faking a victim's voice is not needed, i.e., attackers can simply record their own voice at a close distance and with a high-quality recorder to eliminate background and channel noise in the voice command. Hence, the background and channel noise features can no longer be used to differentiate a fake voice from a real one. Malicious commands are naturally different from the historical records in a VCS, therefore, the approaches in~\cite{108shang2010score,109shang2008preliminary,110shang2008playback} will also fail. The attacker can also synthesize voice commands using a text-to-speech system without the need of voice conversion or copying and pasting and consequently, the approaches in~\cite{119villalba2011preventing,121todisco2016new} will also not work. In summary, the existing techniques built to protect ASV systems are not a good fit for the defense needs of a VCS.

\subsubsection{The Playback Phase}

In ASV applications, the microphone is usually positioned very close to the user (i.e., less than 0.5m). At such distances, some acoustic features can be used to identify the sound source of the speaker, e.g., in~\cite{104shiota2016voice,105shiota2015voice}, the authors use the ``pop noise'' caused by breathing to identify a live speaker. Other efforts~\cite{114korshunov2018use,106li2017study,120witkowski2017audio} do not explicitly use close distance features, but the databases they use to develop their defense strategies were recorded at close distances~\cite{102kinnunen2017asvspoof,107kinnunen2017reddots}, and therefore, these approaches may also implicitly use close-distance features. In contrast, with the help of far-field speech recognition techniques, modern voice controlled systems can typically accept voice commands from rather long distances (i.e., several meters to tens of meters). At such distances, close-distance features cannot be used to distinguish between human speakers and recorded voice, e.g., the pop noise effect quickly disappears over larger distances. 
%As a consequence, close distance acoustic features used to protect ASV may not fit VCS.

\begin{figure}[t]
  \centering
  \includegraphics[width=8.6cm]{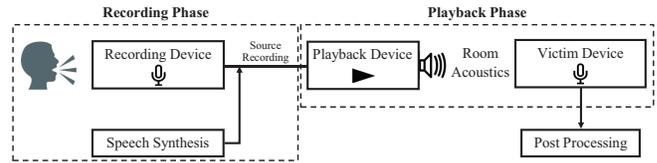}
  \caption{Typical replay attacks include a recording phase and a playback phase. In the recording phase, the attacker records or synthesizes a malicious voice command. In the playback phase, the malicious voice command is transmitted from the playback device to the victim device over the air. Unique aspects of attacks on a VCS (in contrast to an ASV system) are that malicious commands can easily be generated (leaving very few cues in the command itself) during the recording phase and the transmission distances can be very long during the playback phase.}
  \label{fig:replayAttack}
\end{figure}

%\subsection{Sound Source Identification Using Acoustic Cues}
%\label{sec:proposedMethod}

In summary, VCS attack scenarios may leave only very few cues during the recording phase that could help detect a replay attack. Instead, we have to focus on the playback phase, where we have to identify features that discriminate between human and electronic commands, especially when commands are given over larger distances. Modeling the sound production and transmission over long distances with room reverberation is complex, making it difficult to design the required features. Therefore, in this work, we first extract a large acoustic feature set and then use machine learning techniques to identify the discriminative features. Since this is a new direction in protecting attacks on a VCS, existing datasets are difficult to use since they either contain recordings made over short distances~\cite{103delgadoasvspoof,107kinnunen2017reddots} or they contain non-speech content~\cite{113foster2015chime}.
Therefore, we collected our own dataset consisting of voice commands produced by both humans and different playback devices, and recorded at various distances from the speaker in the playback phase (the details of this dataset are described in Section~\ref{sec:data}). We further use the COVAREP~\cite{degottex2014covarep} acoustic feature extraction toolkit, which extracts 74 features per 10ms, and then apply three statistic functions (mean, max, and min) to each feature over the entire voice command sample, which leads to a 222-dimensional feature vector for each voice command sample. We use support vector machine (SVM) with radial basis function (RBF) kernel as the machine learning algorithm.

There are two considerations that facilitate the task of 
building a sound source identification system for VCS using acoustic features. First, the sound source identification can be done in a {\em text-dependent} way, i.e., even though a voice command could be any text, it usually needs to start with a fixed wake word (such as ``Alexa'', ``Hey Google'', or ``Hey Siri''). We therefore only need to identify the source of the fixed wake word, which eliminates the impacts of analyzing different spoken texts.
Second, a VCS typically runs only on some dedicated devices that use a fixed microphone model, e.g., Alexa runs on Amazon Echo devices, while Google Home runs on Google Home devices. This means that the victim device in Figure~\ref{fig:replayAttack} is fixed, which eliminates another variable in the playback phase. Otherwise, different microphones may have different sound collection characteristics (e.g., frequency response), which may be confused with differences in playback device characteristics and thereby affect the identification.

\begin{table*}[]
\centering
\caption{The replay attack range of Amazon Echo Dot and Google Home Mini}
\label{tab:attackRange}
\begin{tabular}{@{}ccccccc@{}}
\toprule
VCS Device       & \multicolumn{3}{c}{Meeting Room} & \multicolumn{3}{c}{Corridor}    \\ \midrule
                 & Headphone  & iPod  & Loudspeaker & Headphone & iPod  & LoudSpeaker \\ \cmidrule(l){2-7} 
Amazon Echo Dot  & 1-19m       & 1-19m  & 1-22m        & 21.0m     & 22.4m & 28.9m       \\
Google Home Mini & 1-18m       & 1-18m  & 1-21m        & 4.4m      & 4.7m  & 21.0m       \\ \bottomrule
\end{tabular}
\end{table*}

\begin{figure}[t]
  \centering
  \includegraphics[width=7cm]{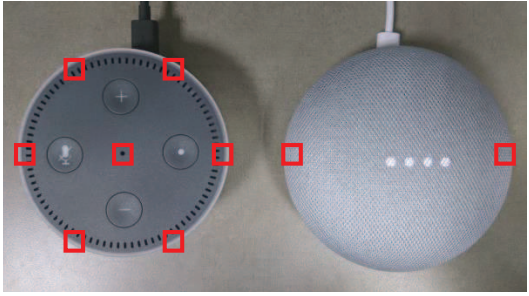}
  \caption{The VCS devices used in our experiments: Amazon Alexa-based Amazon Echo Dot (left) and Google Home Mini (right). The Amazon Echo Dot has 7 microphones, while the Google Home Mini has 2 microphones (the microphone positions are shown with the rectangles). }
  \label{fig:VCS}
\end{figure}

\begin{figure}[t]
  \centering
  \includegraphics[width=7cm]{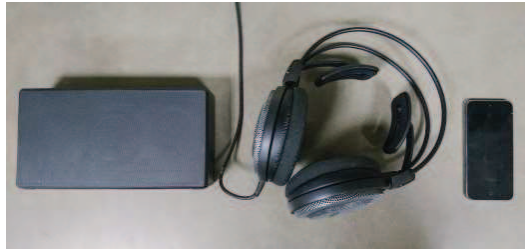}
  \caption{The playback devices used in our experiment: Sony SRSX5 loudspeaker (left), Audio Technica ATH-AD700X headphone (middle), and iPod touch (right). }
  \label{fig:playbackDevice}
\end{figure}

\begin{figure}[t]
  \centering
  \includegraphics[width=7.5cm]{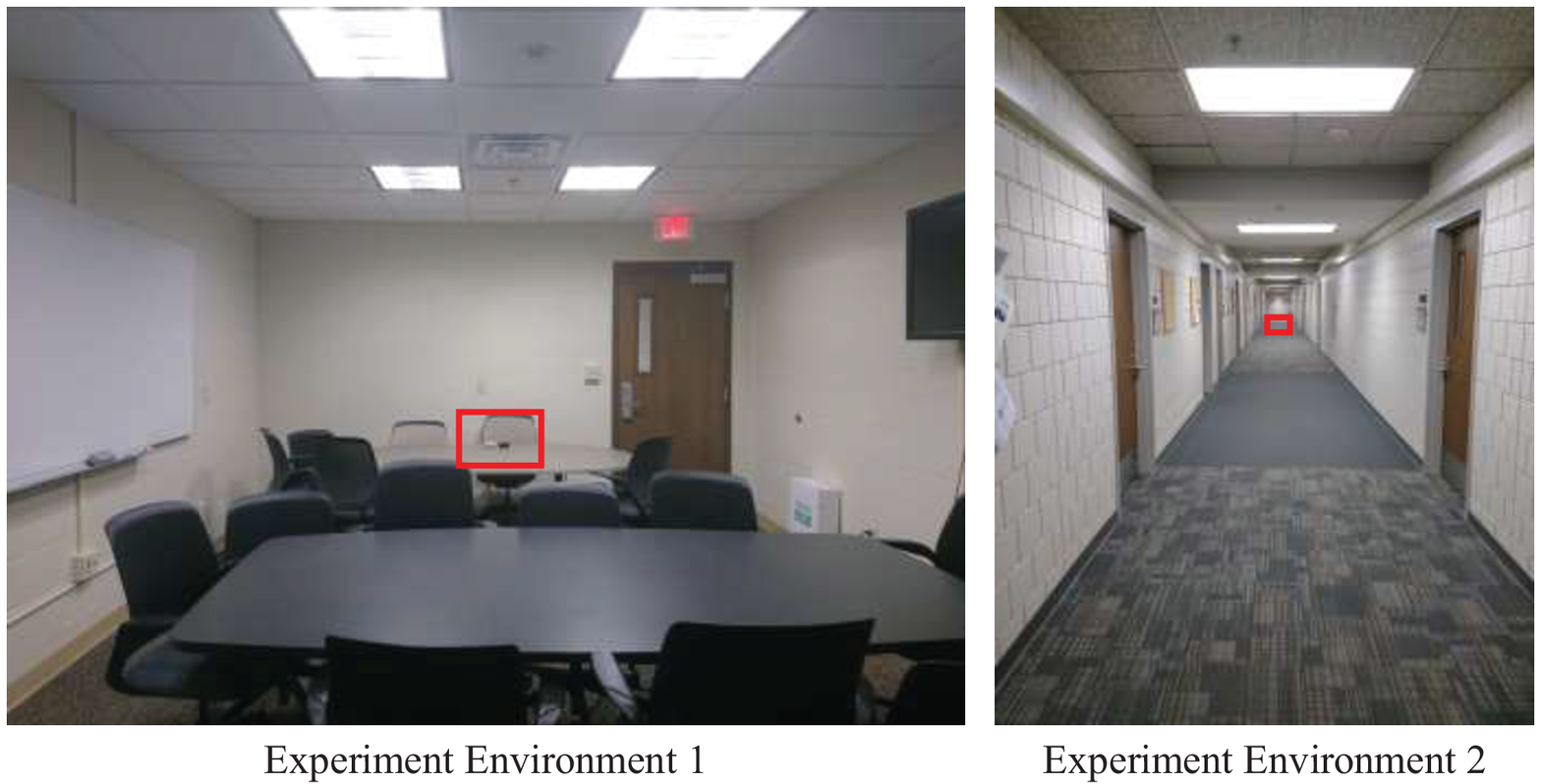}
  \caption{Experiment locations: a typical meeting room (left) and a long corridor (right). The VCS device/microphone location is indicated with the rectangle. The right picture is taken from the edge of the attack range of the Amazon Echo Dot.}
  \label{fig:environment}
\end{figure}

\begin{figure}[t]
  \centering
  \includegraphics[width=7.5cm]{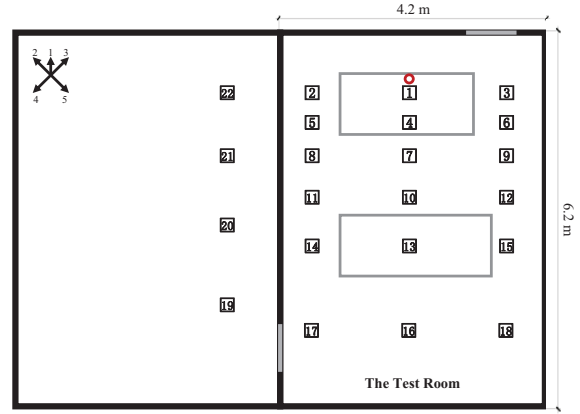}
  \caption{The floor plan of experiment environment 1 (the room to the right). The recording device of the experiments in Section~\ref{sec:data} is placed in the circle. 22 positions are marked with the rectangles (P1-P22). The different directions of the speakers in the experiments in Section~\ref{sec:data} are indicated by the top left arrows.}
  \label{fig:meetingRoom}
\end{figure}

\section{Experimentation}
\label{sec:exp}

\subsection{Replay Attacks on VCSs}

The lack of defense solutions against replay attacks has been reported in multiple previous efforts~\cite{07lei2017insecurity,04alepis2017monkey}, but due to the rapid advances of cloud-based systems, we first evaluate replay attacks with the goal of verifying if state-of-the-art VCS devices will reject replayed voice commands, especially when a sensitive operation is requested.   
%\footnote{The experiments in this section were performed in the United States in April 2018.} rejects replayed voice commands, especially when a sensitive operation is requested. 
We run these experiments with Amazon Alexa and Google Home devices as shown in Figure~\ref{fig:VCS}. The user of these devices is an adult male. 
%Using a high suspicious case, 
The replay attack is performed using a synthetic voice command by a female (implemented using Google Text-to-Speech, i.e., the resulting command will not sound completely natural to humans). The content of the voice command is ``Alexa, buy a laptop'' and ``Hey Google, buy a laptop'' with two subsequent ``Yes'' commands. The voice commands are replayed using a headphone (shown in Figure~\ref{fig:playbackDevice}) at a 50cm distance to the VCS. This attack successfully makes Alexa place an order of a \$350 laptop, while we found that Google Home currently does not allow purchases over \$100. Therefore, we changed the voice command and successfully let Google Home place an order of a \$15 set of paper towels. Further tests also showed that
both devices will still perform the requested action when 
the genuine male voice command and the replayed female voice command are alternated in a single chat. 

Note that both Alexa and Google Home provide a feature to learn the voice of the user (i.e., ``Alexa Your Voice'' and ``Google Home Voice Match''). 
%\textcolor{red}{but they are not used by default.} 
%and mandatory setting. 
In our next experiment, we enable this feature, let the VCS learn the voice of the male user, and then repeat the above described attack. Alexa still accepts the voice command and places the order as before, while Google Home rejects the request, because of the voice mismatch. We then repeat the attack using a pre-recorded voice command of the user and successfully let Google Home place an order. That is, this feature does not provide a strong defense against replay attacks (note that the purpose of the voice learning feature is to provide personalized services rather than addressing a security concern). In addition, this feature also affects the usability of legitimate shared use of a VCS.
%, therefore, \textcolor{red}{most users will keep it disabled. }
%Finally, we also found that not all VCS devices operate in an ``always-on'' mode, which may limit the impact of replay attacks. For example, Apple Siri always listens only when the iPhone is charged, limiting the attack be valid mostly in indoor scenarios (or if a battery case is used outdoor); Google Assistant works in always listening mode only on a few Android devices, e.g., Google Pixel 2.

To conclude, although there are constraints limiting the impact of attacks on VCS devices, state-of-the-art VCS solutions are not able to detect a replayed voice command, which leaves a severe security flaw.

\subsection{Attack Range Analysis}
As discussed in Section~\ref{sec:proposedMethod}, for the proposed defense strategy, we need to identify the source of the voice command within the ``attack range'' of a VCS, i.e., the maximum distance between a playback device and a VCS device. In other words, we need to measure how far away the malicious voice command can come from to still be accepted by the VCS. This attack range depends on three parameters: (1) the playback device itself, which affects the sound production, e.g., sound volume and speaker directivity; (2) the environment, which affects the sound transmission, the background noise, and the room reverberation; (3) the VCS device itself, which affects the sound collection. In this experiment, we test three playback devices: Sony SRSX5 portable speaker, Audio Technica ATH-AD700X headphone, and iPod Touch. The devices are tested in two environments: a typical meeting room and a long corridor, with two VCS devices (Amazon Echo Dot and Google Home Mini), therefore, we have a total of 12 attack conditions. The playback devices and environments are shown in Figures~\ref{fig:playbackDevice},~\ref{fig:environment}, and~\ref{fig:meetingRoom}. 

Since the attacks (with the exception of impersonation) are based on the basic replay attack, in this experiment, we measure the attack range of the basic reply attack. The attack range of its variants may be shorter than this range, e.g., minor perturbations of a machine learning level attack might not be able to be captured by the VCS at larger distances~\cite{12carlini2018audio}. We perform the attacks in different positions in the environment by replaying a synthetic voice command ``Alexa, what's the weather today?'' and ``Hey Google, what's the weather today?'' via the playback devices, repeating the experiments three times. If the VCS accepts any one of the voice commands and reports the weather, we regard the attack as successful. As shown in Table~\ref{tab:attackRange}, all three variables have a large impact on the attack range. In experiment environment 1 (meeting room), the attack range of all three playback devices covers the entire room (P1-P18 in Figure~\ref{fig:meetingRoom}) for Alexa and Google Home. We then opened the door of the room and extended the experiments into the neighboring room (left room in Figure~\ref{fig:meetingRoom}). Here, we find that the attack range can be even larger (P19-P22). We further repeat the experiment in environment 2 (corridor) and measure the longest straight-line attack distance. This experiment provided the following findings:
\begin{enumerate}
\item {\textbf{The attack range is large.}} 
The attack range is larger than expected, e.g., headphones are not designed for replaying sound loudly, but their attack range still covers a typical room. The attack range of a loudspeaker can be over 25m (in the corridor) and go through a wall (in the meeting room). This means that attacks can be performed over long distances and even from a different room.
\item {\textbf{The attack range depends on the VCS device.}}
The Amazon Echo Dot and the Google Home Mini differ in their abilities to capture sound. That is, the Amazon Echo Dot picks up commands over longer distances and can therefore be attacked over a substantially larger range. This is likely due to the fact that the Amazon Echo Dot has seven microphones, while the Google Home Mini has only two. Such a microphone array is beneficial for far-field speech recognition, but while this improves the usability of a VCS, it also increases the risk of being attacked.  
\item{\textbf{The attack range depends on the environment.}}
The attack range is also determined by the environment,  e.g., the straight-line attack distance of headphone/iPod to Google Home Mini in the corridor is only around 4.5m, but their straight-line attack distance in the meeting room is larger than 6m. This is likely due to the fact that sound waves are attenuated differently in an open space (corridor) compared to a closed space (meeting room). 
\item{\textbf{The attack range depends on the playback device.}} While it is unsurprising that the loudspeaker has a greater volume and therefore a larger attack range, the relationship between attack range and type of playback device is more complex than simply a difference in volume. 
For example, the attack range difference between headphone and Amazon Echo Dot and between headphone and Google Home Mini is much larger than that of the loudspeaker, indicating that there are other factors at play besides a volume difference. The goal of this work is to exploit such differences to help identify the sound source.
\end{enumerate}

\subsection{Sound Source Identification}
\label{sec:data}

%Based on these experimental measurements in the last experiment, and considering making the experiment close to practical setting, this experiment is conducted in the above mentioned meeting room (experiment environment 1). 

Considering that VCS devices are more likely to be placed in a typical room rather than a corridor, we limit the following experiments to the meeting room environment (experiment environment 1). We use an iPod Touch to record the sound, simulating the sound collection of a VCS device, because we cannot directly access the sound stored in a commercial VCS. We refer to this iPod Touch as ``sound collector'' in the remainder of this section. The sound collector is placed on one side of the room. We then record voice command samples ``Alexa'' at 22 different positions, i.e., P1-P22 (shown in Figure~\ref{fig:meetingRoom}). The voice command is produced by (1) a loudspeaker, (2) an iPod, (3) a headphone, and (4) a human speaker. The voice commands (which are then replayed by the playback devices) are recorded from a human speaker using a professional recorder (Tascam DR-05) at a close distance and in a quiet environment. This emulates a realistic attack scenario, i.e., this will leave as few cues in the recording phase as possible (the voice command itself is almost exactly the same). At each position, we also record multiple voice command samples that are produced when the speaker faces different directions. In more detail, for the iPod and headphone, we record voice commands in positions P1-P18 and at each position, we record three directions (all in forward directions to the sound collector, i.e., directions 1-3 shown in Figure~\ref{fig:meetingRoom}). For the loudspeaker and human speaker, we produce voice commands in positions P1-P22 and at each position, we use five directions (both forward and backward directions relative to the recorder, i.e., all directions 1-5 shown in Figure~\ref{fig:meetingRoom}). This is done because our previous results showed that the loudspeaker has a larger attack range. Overall, after eliminating a small number of samples due to undesired noise, we obtained a dataset consisting of a total number of 296 voice command samples produced by different sound sources (loudspeaker: 106, iPod: 44, headphone: 46, human: 100).

We then normalized each voice command sample waveform to have the same max amplitude of 1 in order to make all voice command samples have the same volume. We then extract a 222-dimensional feature vector using the COVAREP toolkit and feed it to the SVM with an RBF kernel as described in Section~\ref{sec:proposedMethod}. We use empirical hyperparameters: the cost of SVM $C$ = 1 and the $\gamma$ of the RBF kernel = 0.25 (the reciprocal of the number of classes). We then evaluate the results using a ten-fold cross validation, where the training and testing sets are independent from each other in each round. 

\begin{figure}[t]
  \centering
  \includegraphics[width=6.5cm]{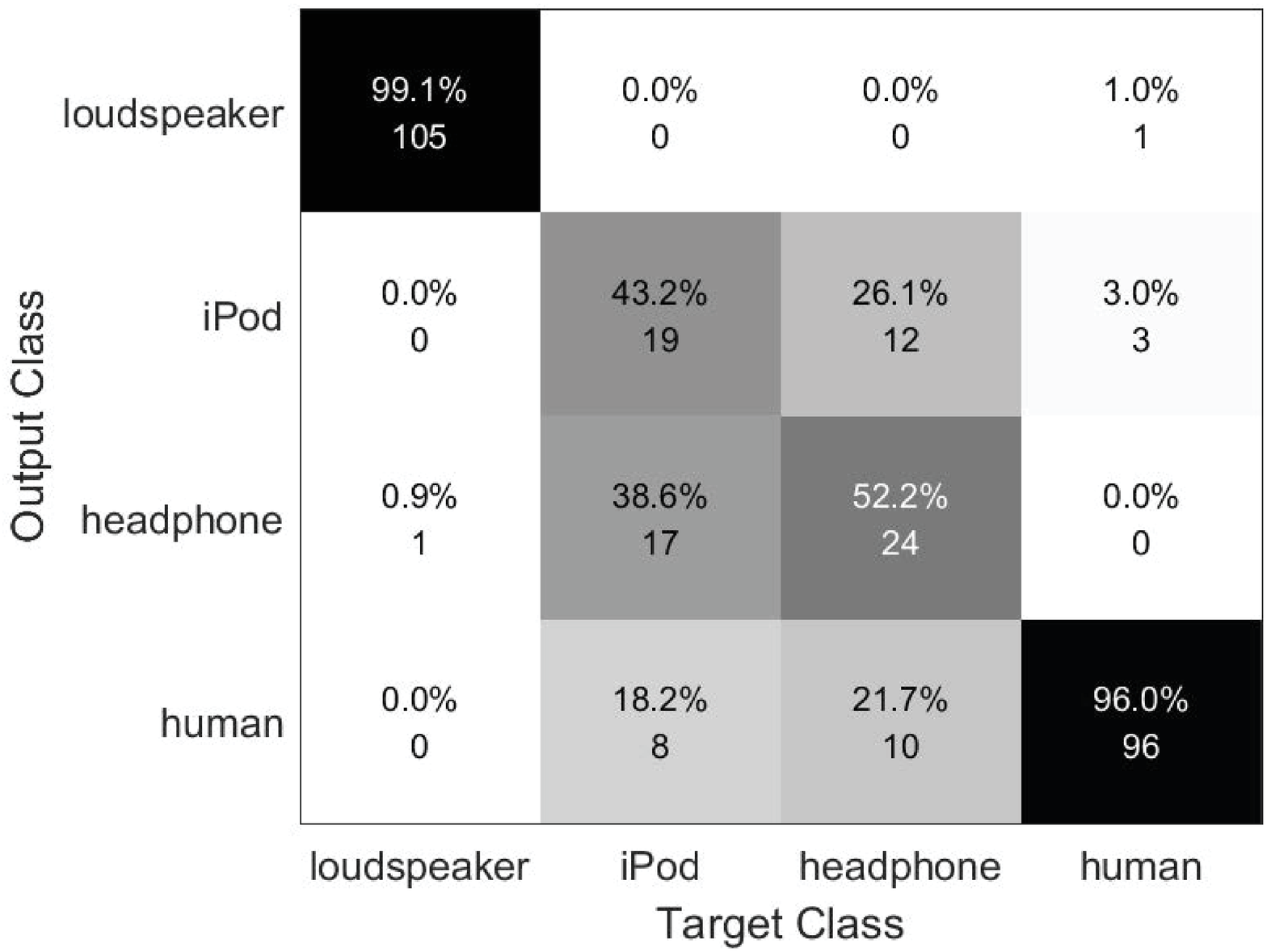}
  \caption{The confusion matrix of the sound source classification test.}
  \label{fig:confMat}
\end{figure}

The results of the experiment are shown in Figure~\ref{fig:confMat}. For the human speaker-playback devices classification, i.e., where we consider loudspeaker, iPod, and headphone as a single class, we achieve an F1-score of 0.94 on the playback device and 0.90 on the human speaker class, leading to an average F1-score of \textbf{0.92}. We further find that human voice commands are more likely to be misclassified as those reproduced by the iPod and the headphone. That is, if we regard iPod and headphone as a single class and take out the loudspeaker class (since it is rarely misclassified with other classes), the F1-score is 0.87 in the new iPod-headphone class and 0.90 in the human speaker class, leading to an average F1-score of \textbf{0.89}. This result tells us that the loudspeaker is very different from the human speaker and even other playback devices, which can therefore be easily identified. The iPod and headphone devices are closer to the human speaker, but can still be effectively identified. Interestingly, the microphone and iPod are difficult to be classified as each other. All these results verify the feasibility and effectiveness of using acoustic cues to identify the sound source of a voice command.

\begin{table}[t]
\centering
\caption{Selected Acoustic Features by Correlation-based Feature Subset Selection}
\label{tab:feature}
\begin{tabular}{@{}p{1.98cm}p{6.4cm}@{}}
\toprule
Statistic Function    & \multicolumn{1}{c}{Feature Selected (28)}         \\ \midrule
Max                   & Fundamental Frequency, MFCC(3,4,8,10,13)          \\
Min                   & MFCC(1,3,5,7,10,22)                               \\
\multirow{2}{*}{Mean} & Harmonic Structure (H1H2), HMPDM(4,6), HMPDD(0)            \\
                      & MFCC(1,3,4,7,8,12,13,14,15), Peak Slope  \\ \bottomrule
\end{tabular}
\end{table}

We further use a feature selection algorithm (which we did not use for learning the model, because it could lead to overfitting in a small dataset) to analyze the discriminative features. The results are shown in Table~\ref{tab:feature}, where we provide the combination of features that contribute to the classifier, including fundamental frequency, Mel-cepstral coefficients (MFCCs), and harmonic model and phase distortion mean and deviation (HMPDM, HMPDD). This indicates that modeling such a classifier is complex and that the use of a machine learning model is essential. 

Finally, it needs to be mentioned that while the empirical results are encouraging, the experiments are performed in fixed environments using three representative playback devices only. Therefore, the learned model may lack generalization. In practice, there are infinite options for playback devices and environments. Further, speaker variability should also be considered. As a consequence, an important future step will be to build a larger database containing a variety of different conditions, which can then serve as the basis for the development of more generalized machine learning models.

\section{Conclusions}
\label{sec:conclude}
In this work, we first review state-of-the-art attack technologies and find that all of them (with the exception of the impersonation attack) are based on the replay attack, where the malicious voice command is produced by a playback device. Based on the fact that legitimate voice commands should only come from a human speaker, we then proposed a novel defense strategy that uses the acoustic features of a speech signal to identify the sound source of the voice command and only accept the ones coming from the human. Compared to existing defense strategies, the proposed approach has the advantage that it minimally affects the usability of the VCS, while being robust to most types of attacks. Since identifying the sound source of voice commands in a far-field condition has barely been studied before, we first measure the practical attack ranges of modern VCS devices (i.e., Amazon Alexa and Google Home) and then use the results to construct a dataset consisting of both genuine and replayed voice command samples. We then use this dataset to develop a machine learning model that can be used to distinguish the human speaker from the playback devices. Finally, our proof-of-concept experiments verify the feasibility of the proposed approach.

\bibliographystyle{IEEEtran}
\bibliography{reference.bib}

\end{document}